\documentclass[11pt,eqs]{article}

\usepackage{mathtools}
\usepackage{xcolor}
\usepackage[margin=1in]{geometry}
\usepackage{latexsym,amsmath,amssymb,amsthm,amstext,tikz}
\usepackage{fancyhdr}
\usepackage{graphicx,relsize}
\usepackage{multirow,multicol}
\usepackage{pdflscape}

\def\cleq{\setcounter{equation}{0}}

\DeclarePairedDelimiter\bra{\langle}{\rvert}
\DeclarePairedDelimiter\ket{\lvert}{\rangle}
\DeclarePairedDelimiterX\braket[2]{\langle}{\rangle}{#1 \delimsize\vert #2}
\DeclareMathOperator{\Tr}{Tr}

\title{Normal ordering in string theory and the canonical phase space quantization
\thanks{Work supported in part by The University of Belgrade, Belgrade, Serbia and by the home institutions of the authors.}}
\author{  Ljubica D. Davidovi\'c \thanks{e-mail: ljubica@ipb.ac.rs},
Milo\v{s} D. Davidovi\'{c} \thanks{e-mail: davidovic@vin.bg.ac.rs},
and Milena D. Davidovi\'{c} \thanks{e-mail: milena@grf.bg.ac.rs}\\
{\it  Institute of Physics,  National Institute of the Republic of Serbia,}\\
{\it   University of Belgrade},\\
{\it Vin\v ca Institute of Nuclear Sciences, National Institute of the Republic of Serbia,}\\
{\it   University of Belgrade, and}\\
{\it Faculty of Civil Engineering, University of Belgrade}}

\begin{document}
\maketitle

\begin{abstract}

We consider quantization techniques with a particular emphasis on the  phase space quantization, as encountered in quasi-probabilistic descriptions and in string theory. We investigate what is lost by adopting the normal ordering prescription, by revisiting fundamental quantization concepts and the correspondence rules that define the well-known quasi-distributions: the Wigner, Husimi, and P functions —
some of which are inherently ordering-dependent.
The actual quantum state dictates 
the necessity of employing discrete countable structures or, alternatively, the complex phase space formulation based on coherent states,
and the quasi-distributions depict the existence of the phase space dynamics, of the wave translations and the propagating directions.

\end{abstract}

\section{Introduction}

All quantization procedures share a common foundation: the Dirac quantization rule, which replaces the classical Poisson bracket between coordinate and momentum with the corresponding quantum commutator
\begin{equation}\notag
\{x^\mu(\sigma),\pi_\nu(\bar\sigma)\}=\delta^\mu_\nu\delta(\sigma-\bar\sigma)\longrightarrow[\hat{q},\hat{p}]=i\hbar\hat{1}.
\end{equation}
This prescription arises from the passage from the configuration space to the phase space, and from the Lagrangian formulation of dynamics to the canonical Hamiltonian, connected through the definition of the canonical momentum and the constraints it obeys.
The classical phase space $(q,p)$, being a well measurable
property of a physical object is replaced by
 the non-commuting operators $\hat{q},\hat{p}$ acting  on  the physical states,
occupied only with  a certain probability.
 Operators corresponding to  incompatible observables cannot be measured simultaneously, and the minimum value of the uncertainty relations, evaluated for the considered quantum state represents the best achievable precision in their joint measurement.
The connection between  other classical and quantum variables  is established through correspondence rules.

 The quantization rules are fundamentally based on the choice of normal coordinates that associate the position with the complete set of basis functions. In most cases, these are chosen to be plane waves, which differ from one another only by a phase factor.
Analogously, in the Weyl quantization prescription, the connection between the classical and quantum variables  is established through
the generalized 
Fourier transform for both phase space coordinates.
It is supplemented by its
closure properties, dictating the precise form of the Dirac delta.  The Weyl quantization is symmetric: it treats $\hat{ q}$ and $\hat{ p}$ on the equal footing by averaging over all possible orderings with equal weight. This makes it particularly natural for the phase-space formulations of the quantum mechanics.

In the standard formulation of quantum mechanics one typically adopts a representation in which either the quantum state or the quantum operator is stationary.
The transition between these pictures is governed by the unitary evolution operator.
In the phase-space formulation of quantum mechanics the time exists only through its derivative given by a Poisson bracket algebra with the corresponding Hamiltonian.
This is a canonical description,
potentially 
the best setting for 
understanding the connections between the classical and quantum mechanics, and the relations between the 
classical and quantum states of nature. 
The standard quantum states exist as projections of the density matrix,
where the density operator represents the physical state.
The ordering problem,
the multiple choice of quantum operators corresponding to one classical function, is thoroughly investigated in this formulation, relating Wigner, Husimi and Glauber-Sudarshan quasi-distributions to
symmetrical, anti-normal and normal orderings.

On the contrary to the theoretical connections,
the real, physical connection between the quantum and the classical theories is established 
through the quantum states of light probing the media,
where their resulting physical states give incite into the mutual type of interaction through the obtained statistics.

The significant role in describing the classical to quantum variables transition,
together with the connection between the vacuum and coherent states and also
the phase-space transformations (doubled translation, shift operator) has 
 the Weyl's displacement operator

\begin{equation}
 D(q,p)=e^{-\frac{i}{\hbar}(q\hat{p}-p\hat{q})}=e^{\alpha \hat{a}^{+}-\alpha^\ast \hat{a}}=D(\alpha,\alpha^\ast),\notag
\end{equation}
which is itself invariant to the transition from the real to the complex phase space.
All descriptions are related to these five operators.
The exponentials of creation and annihilation operators
are known to produce convergence for the normal ordering and divergence for the anti-normal ordering.
The related problem is whether one should use 
the ordinary expanding of the exponential, or the expansion formulas classified by the 
commutator properties of the non-commuting operators.

In measurements, the connection between the real and the complex phase space variables used for description of light is again established  through the normal coordinates,
by splitting the laser beam leaving one of the parts free of interaction and then again combining them the at phase component appearing as amplification is conjugated to 
position and cosine function while its sine counterpart, $\frac{\pi}{2}$ delayed is the quadrature component.

For the  algebraical part of quantization rules, the standard  Poisson algebra is promoted to the commutator algebra which  inherits
the structure by preserving the trigonometry relations. As for the states, the complex phase space operators introduce the number states which do not produce  a good classical limit, but whose ground state, representing the absence of particles is the most important in investigating the quantum fluctuations described by the quantum field theory. For the existence of the mean trajectory, one needs the superposed state, the coherent state.
The phase space quantum mechanics additionally
distinguishes between the pure and mixed quantum mechanical states. The pure state cannot
become a mixed state without the interaction with it's environment and the real quantum systems are never isolated.

\section{Bosonic string theory}
\cleq

It is interesting, that in the  string theory quantization, the normal ordering is demanded and it here reads
put smaller indexed to the left. 
However, the awareness of the ordering arbitrariness is noted by a undetermined constant,
which is afterward assigned to fulfill the dimensionality condition.
The idea behind the string theory is of the fundamental string 
propagating and  forming the world-sheet,
the form of which is governed by the space-time of the higher dimension the string exists in.
The mathematical problem of embedding the two-dimensional world-sheet $\Sigma$  into the space-time,
is done on the several levels:
the coordinate settings, the distance measurement relations and finally by the exact form of the theories describing the propagation.

The bosonic string dynamics  can be described   in the flat space-time by
two actions:
the Nambu-Goto and the Polyakov actions, the first being proportional to the world-sheet area
\begin{equation}\label{eq:ng}
S=\kappa\int_{\Sigma} d^{2}\xi\,\sqrt{-\det \{\partial_\alpha x^\mu
\partial_\beta x^\nu G_{\mu\nu}\}}\,,
\end{equation}
 and the second 
being its Lagandre transformation for the induced metric $\partial_\alpha x^\mu
\partial_\beta x^\nu G_{\mu\nu}$,
measuring the square distance between points on the world-sheet $\xi^\alpha$ u $\xi^\alpha+d\xi^\alpha$,
corresponding to the shift of coordinates 
$x^\mu\rightarrow x^\mu+\partial_\alpha x^\mu d\xi^\alpha$.

Non-linearity is what makes the first action difficult to deal with, so one adds the metric on the world-sheet by
\begin{equation}\label{eq:nglag}
S=\kappa\int_{\Sigma} d^{2}\xi\,
\bigg{[}
\sqrt{-\det \{
g_{\alpha\beta}\}}
+\lambda^{\alpha\beta}(g_{\alpha\beta}-
\partial_\alpha x^\mu\partial_\beta x^\nu G_{\mu\nu})
\bigg{]}.
\end{equation}
This action incorporates both Nambu-Goto and Polyakov actions because of the variation principle,
for the equation of motion following from the variation over the Lagrange multiplier $\lambda^{\alpha\beta}$ one regains the NG-action, while for the equation of motion obtainable varying over the metric $g_{\alpha\beta}$
one obtains the Polyakov action
\begin{equation}\label{eq:pol}
S
=\frac{\kappa}{2}\int_{\Sigma}d^{2}\xi
\sqrt{-g}\,g^{\alpha\beta}
\partial_\alpha x^\mu
\partial_\beta x^\nu G_{\mu\nu}.
\end{equation}

The Polyakov action itself becomes the NG-action for the equation of motion, obtained varying by $g_{\alpha\beta}$
\begin{equation}\label{eq:emP}
T_{\alpha\beta}=0,
\end{equation}
where the energy-momentum tensor is preserved due to the re-parametrisation invariance of the action
\begin{equation}\label{eq:ri}
\nabla^\alpha T_{\alpha\beta}=0.
\end{equation}

It is interesting that the Nambu-Goto action which is explicitly
\begin{equation}
S_{NG}=
\kappa\int_{\Sigma}d^{2}\xi\sqrt{-\gamma}\,,\quad
\gamma_{\alpha\beta}=G_{\mu\nu}
\partial_\alpha x^\mu\partial_\beta x^\nu,
\quad\gamma=(\dot x)^{2}x^{\prime 2}-(\dot{x}\cdot x^\prime)^{2},
\end{equation}
has a zero canonical Hamiltonian density.
The canonical momentum is
\begin{equation}
\pi_\mu=\frac{\partial{\cal L}}{\partial\dot{x}^\mu}=-\kappa(-\gamma)^{-1/2}\Big{[}
\dot{x}_\mu x^{\prime 2}-(\dot{x}x^\prime)x^\prime_\mu
\Big{]},
\end{equation}
with the two primary constraints
\begin{equation}\label{eq:PC}
\pi\cdot x^\prime=0,\quad \pi^{2}+\kappa^{2}x^{\prime 2}=0,
\end{equation}
hence the canonical Hamiltonian is just  
\begin{equation}
{\cal H}=\pi\cdot\dot{x}-{\cal L}=0,
\end{equation}
and the dynamics is determined solely by the constraints.
The same constraints exist for the Polyakov action as well, but for (\ref{eq:emP}).

Once the measures of the space-time and the world-sheet are coordinated, one can partially fix the parametrization on the world-sheet by taking the conformal (light-cone, orthonormal) gauge.
In the light-cone parametrization
\begin{equation}
 \xi^\pm=\tau\pm\sigma\;\text{and}\;\partial_\pm=\frac{1}{2}\big(\partial_{0}\pm\partial_{1}\big),
\end{equation}
and in the light-cone gauge where the re-parametrization invariance and the Weyl re-scaling are used to set the Minkovski metric on the world-sheet, the Polyakov action becomes
\begin{equation}
S=\frac{\kappa}{2}\int\,d^{2}\xi\,\eta^{\alpha\beta}\partial_{\alpha}x^\mu\partial_{\beta}x_\mu=-2\kappa
\int\,d^{2}\xi\,\partial_{+}x^\mu\partial_{-}x_\mu,
\end{equation}
with the equation of motion $\partial_{+}\partial_{-}x_\mu=0$, the canonical momentum $\pi_\mu=-\kappa\dot{x}_\mu$,
the boundary conditions $\gamma^{0}_\mu\delta x^\mu=0$ with $\gamma^{0}_\mu=\frac{\partial{\cal L}}{\delta x^{\prime\mu}}=\kappa x_\mu^\prime$
and the Hamiltonian
\begin{equation}
H=-\kappa\int_{0}^{l}d\sigma\big{(}(\partial_{+}x)^{2}+ (\partial_{-}x)^{2}\big{)}.
\end{equation}
The constraints (\ref{eq:PC}) together with the light-cone gauge momentum give
\begin{equation}
 \dot{x}\cdot x^\prime=0,\quad(\partial_{0}x)^{2}+(\partial_{1}x)^{2}=0,
\end{equation}
which can be combined into 
\begin{equation}\label{eq:constraints}
(\dot{x}\pm x^\prime)^{2}=0.
\end{equation}
The equations (\ref{eq:emP}) and (\ref{eq:ri}) become
\begin{eqnarray}
&&T_{++}=(\partial_{+}x)^{2}=0,\quad \partial_{-}T_{++}=0,
\nonumber\\
&&T_{--}=(\partial_{-}x)^{2}=0,\quad\partial_{+}T_{--}=0.
\end{eqnarray}

This is where the quantum string theory in the light-cone quantization procedure is based.
On the conserved constraints for the consistent  distance measures.

\subsection{The classical Virasoro generators}

Depending on the boundary conditions the string coordinates should obey,
one chooses different, the most convenient normal coordinates.
For the closed string $x^\mu(\sigma)=x^\mu(\sigma+2\pi)$ one expands the coordinate into the Fourier series over $e^{in\sigma}$, while for the open string satisfying the Neumann
boundary condition 
\begin{equation}
x^{\prime\mu}\Big{|}_{0}^{\pi}=0,
\end{equation}
one chooses $\cos(n\sigma)$.
The appropriate expansion is then used to solve the equation of motion.
The solution defining parameters by parts inherit the standard Poisson bracket between the 
coordinates and momenta.

Consequently, 
the constraints (\ref{eq:constraints}) obtain their expansion to normal coordinates what defines the Virasoro operators $L_{n}$. For the closed string there are two constraints and therefore
\begin{eqnarray}
&&L_{n}={\kappa}\int_{-\pi}^{\pi}d\sigma\,e^{in\sigma}T_{++},
\nonumber\\
&&\bar{L}_{n}={\kappa}\int_{-\pi}^{\pi}d\sigma\,e^{in\sigma}T_{--}.
\end{eqnarray}
For the open string one extends the domain from $(0,\pi)$ to $(-\pi,\pi)$ with
$x^\mu(\tau,-\sigma)=x^\mu(\tau,\sigma)$ to obtain
\begin{equation}
\partial_{+}x^\mu(\tau,\sigma)=
\partial_{-}x^\mu(\tau,-\sigma),
\end{equation}
and is therefore left with only one constraint over the extended domain
$(\partial_{+}x)^{2}=0;\sigma\in(-\pi,\pi)$.
Hence, the
Virasoro generators are
\begin{equation}
L_{n}={\kappa}\int_{-\pi}^{\pi}d\sigma\,e^{in\sigma}T_{++}.
\end{equation} 
Their building blocks $\sim\partial_\pm x^\mu$ and $\sim\partial_{+} x^\mu$  Fourier transforms are the objects which will be promoted to the creation and annihilation operators in the quantum theory.

In the  light-cone quantization, the classical Virasoro generators are promoted to the Virasoro operators for the transverse coordinates.
The light-cone coordinates are defined 
by choosing two out of $d$, $x^\mu$ directions, with
$x^\pm=\frac{1}{\sqrt{2}}\big(x^{0}\pm x^{1}\big)$, naming the rest of $(d-2)$ directions  $x^{I}$ the transverse coordinates,
making  the overall change of coordinates $x^{\mu}\rightarrow(x^{+},x^{-},x^{I})$.
The creation operators are set to be the negative frequency modes of the transverse oscillators $\alpha^{I}_{-n}=a^{+}_{n}\sqrt{n}$.
The normal ordering in string theory therefore reads put smaller indexed to the left, which is considering the commutation relations significant only for pairs $n,-n$.
Consequently,  only the operator $L_{0}$ may be changed by a constant reflecting the arbitrariness of the ordering prescription \cite{BBS}.
That is, only this operator is not completely determined by its classical expression \cite{BLT},
and it is relevant for the mass spectrum.

Fixing the world-sheet  residual invariance, by connecting one of the light-cone coordinates to the time parameter $\tau$, the other coordinate can  for the constraints be 
completely determined by the transverse coordinates. Hence, the dynamics is solely determinable by the transverse coordinates
\begin{equation}
S=\frac{\kappa}{2}\int\,d^{2}\xi\,\eta^{\alpha\beta}\partial_{\alpha}x^{I}\partial_{\beta}x_{I}.
\end{equation}
For the open string satisfying the Neumann boundary conditions
$\kappa x_I^\prime\Big{|}_{\sigma=0}^{\sigma=\pi}=0$
one has
\begin{equation}
x_I=x_I(\tau)+\sqrt{\frac{2}{\pi\kappa}}\sum_{n=1}^{\infty}\frac{q^{I}_{n}(\tau)}{\sqrt{n}}\cos(n\sigma),
\end{equation}
and the action hence describes  infinitely many linear harmonic oscillators with the discrete frequency $n$ \cite{BZ}
\begin{equation}
S=\int\,d\tau\,\Big{[}\frac{\kappa\pi}{2}
\dot{x}^{I}(\tau)\dot{x}_{I}(\tau)
+\sum_{n=1}^{\infty}\Big{(}
\frac{1}{2n}\dot{q}^{I}_{n}(\tau)\dot{q}^{I}_{n}(\tau)
-\frac{n}{2}q^{I}_{n}(\tau)q^{I}_{n}(\tau)
\Big{)}\Big{]}.
\end{equation}

In consideration of the string propagating  in both metric and Kalb-Ramond fields
$\Pi_{\pm\mu\nu}=
B_{\mu\nu}\pm\frac{1}{2}G_{\mu\nu}$,
in the Dirac treatment of the boundary conditions, their conservation in time is provided.
It is interesting that,
instead of the pure normal coordinates, one introduces the  effective variables as the solution of the $\sigma$-dependent constraints defined as 
power series of the secondary constraints (the initial being the boundary condition).
Hence, the Poisson bracket of the phase-space variables is replaced by the Dirac bracket between the effective phase space variables. In quantization their ordering matters \cite{DSJHEP2011}. 
In particular, for the mixed Neumann and Dirihlet boundary conditions
\begin{eqnarray}\label{eq:bccan}
_{_{N}}\!\gamma^{0}_{a}&=&
\Pi_{+ab}(G^{-1})^{bc}j_{-c}
+\Pi_{-ab}(G^{-1})^{bc}j_{+c},
\nonumber\\
_{_{D}}\!\gamma^{i}_{0}&=&\kappa\dot{x}^{i}=\frac{1}{2}(G^{-1})^{ij}(j_{+j}+j_{-j}),
\end{eqnarray}
with $j_{\pm\mu}=\kappa G_{\mu\nu}\partial_\pm x^\nu$,
the Dirac consistency procedure yields  the existence of the effective variables
\begin{equation}\label{eq:resenjex}
x^{\prime\mu}=
\begin{cases}
q^{\prime a}-\theta^{ab}p_{b}\\
\bar{q}^{\prime i},
\end{cases}
\quad
\pi_\mu=
\begin{cases}
p_{a},& {\mathsmaller{\mu=a}},\\
\bar{p}_{i}-2\kappa B_{ij}\bar{q}^{\prime j},& {\mathsmaller{\mu=i}}.
\end{cases}
\end{equation}

\section{Quasi-distributions}
\cleq

The quasi-probability distributions represent a classical object
best describing the quantum state, being
described by a density matrix.
The main features of the two quasi-distributions representing two quantization directions are the
preservation of  the  Galilei transformations
acting on the wave function for the Wigner function and the 
mandatory use of the superposed state enabling the mean trajectory
for the  the Q-quasi-distribution  introduced by Husimi.

The main difference between the quasi-probabilistic description and the standard Schrodinger quantum mechanics
is in representing the quantum state.  The first connects to the second,
by the projections onto the states of the standard 
description.
Both quantum descriptions connect to the classical theory by the mean value equations \cite{HOSW}. Explicitly,
the correspondence rules in terms of quasi-distributions are obtained, once the integral representation of the operator expectation value is set.

The quantum mechanics in the phase space formalism,
connects every quantum mechanical operator to a function on the phase space,
the so called operator`s symbol,
 through an integral formula
for their expectation value,
defined for the quasi-probability at hand
\begin{equation}
\langle\hat{O}\rangle=\int\!\!\!\int\,dadb\,
f(\hat{O})(a,b)g(\hat{\rho})(a,b).\notag
\end{equation}
The integration goes over the eigenvalues of the two basical operators.

 The Wigner function was introduced to describe the general quantum states in the real phase space, whereas the Husimi and Glauber–Sudarshan distributions are inherently tied to coherent states and the complex phase space. Nevertheless, when expressed in the complex representation, all three quasi-probability distributions correspond to distinct operator orderings — symmetric, anti-normal, and normal, respectively. This observation once fostered the idea that the operator-ordering ambiguity could be resolved imployingthe  linear interpolations of the quasi-distributions \cite{AW}.
Then, all three quasi-distributions can be represented by a $s$-parametrized distribution $F(\alpha,s)$
on a complex phase space (characterized by $a,a^{+}$) \cite{JBM},
given by
\begin{equation}
F(\alpha,s)=\frac{1}{\pi^{2}}\int\,d^{2}\beta\,
G(\alpha,s)
e^{\alpha\beta^\ast-\alpha^\ast\beta},
\end{equation}
with $G$ being the generalized characteristic function
$
G(\alpha,s)=\Tr\Big{(}
\hat{D}(\beta)\hat{\rho}
\Big{)}e^{\frac{s}{2}|\beta|^{2}}.
$
In fact,  if the commutation relation is inherited from the quantization procedure,
its incorporation into the theory at hand can only be through the consideration of all
orderings, inseparably.

The linear harmonic oscillator quantization is dictating the connection 
between the quasi-distributions.
This description is incorporating the 
coordinate and momentum representations together with
the eigen states of the occupation number operator
and the coherent states.
The creation and annihilation operators come with a property of 
being applicable to both of these  states,
and the latter give the Poissonian distribution of photons and the Gaussian distribution of quadratures. The explicit action of the operators resembles counting and re-scaling. 
The examination of the exact statistics  reveals the border of the particle on the path description to the wave particle duality,
including the wave interference.

The alternative to the operator formalism
is a deformation quantization.
Instead of operator multiplication one deals with the multiplication of functions,
where the multiplication law is  deformed in order to fulfill the
corresponding quantization scheme.
In quantum phase space foundations, this corresponds to the star product \cite{MS} acting between the operator symbols.

Based on the same equation for the expectation value
\begin{equation}
\langle\hat{O}\rangle=\int\!\!\!\int\!\!\!\int\,dxdy dy^\prime\,
\psi^\ast(y)\psi(y^\prime)\delta(y-x)\hat{O}\delta(y^\prime-x),\notag
\end{equation}
but considering the 
 resolution of the Dirac
delta product existing in the  integrand,
there exists the 
 large class of quasi-distributions defined by Cohen.
It is interesting that in string theory, the representation of the delta function \cite{LD}
acting on two  string's space parameters causes the change in algebroidal properties,
including the form of the multiplication between the parameters labeling the generators of the allowed
background transformations.


\section{Wigner function}
\cleq

If the quantum theory is to be compared to the classical theory,
because of its predictions having only a certain probability,  due to the non-commuting operators of coordinates and momenta used to define it,
one should align it with a the classical statistical theory.
However, the classical properties of the distribution functions are lost.
E. Wigner introduced the quasi-probability distribution function to describe the quantum corrections, representing the fact that although the projectors to states with determined values of coordinate and momentum are known,
$\ket{q}\bra{q}$ and $\ket{p}\bra{p}$, the same is not true for the phase space coordinates $(q,p)$.

There are two fundamentally different approaches in defining the Wigner function.
The first offers a simple physical interpretation, placing the definition of  the Wigner function into the linking of the
momentum and coordinate representations of the physical state. This approach, is based on the Schrodinger equation solution for the free particle.
The derivation can be found in \cite{Manko.First}.

For a discrete spectrum of a particle, the density matrix 
\begin{equation}\label{eq:den}
\hat\rho=\sum_{n}w_{n}\ket{\varphi_{n}}\bra{\varphi_{n}}
\end{equation}
has the coordinate and momentum representations given by
\begin{equation}
\rho(q,q^\prime)=
\bra{q}
\hat\rho
\ket{q^\prime}=
\sum_{n}w_{n}\varphi_{n}(q)\varphi^\ast_{n}(q^\prime),
\end{equation}
\begin{equation}
\rho(p,p^\prime)=
\bra{p}
\hat\rho
\ket{p^\prime}=
\sum_{n}w_{n}\varphi_{n}(p)\varphi^\ast_{n}(p^\prime).
\end{equation}
Inserting  the identities $\hat{1}_{q}=\int dq\ket{q}\bra{q}$,
into the diagonal momentum representation, one obtains
\begin{equation}
\rho(p,p)
=\frac{1}{(2\pi\hbar)^{1/2}}\int dqdq^\prime\,\rho(q,q^\prime)\,
e^{\frac{i}{\hbar}p(q^\prime-q)},
\end{equation}
with the plane wave
$\braket{q}{p}=\psi_{p}(q)=A e^{i\hbar^{-1}pq}$,
 being normalized to a Dirac delta function, implying
\begin{equation}
\psi_{p}(q)=\frac{1}{(2\pi\hbar)^{1/2}}e^{i\hbar^{-1}pq}.
\end{equation}
Then, re-scaling the coordinates, without losing the generality and coupling the Jacobiator to the 
coordinate not connected to the momentum, one obtains the definition of the Wigner function
\begin{equation}\label{eq:Wig}
W(p,q)
=\int du\,
\rho\big(q+\frac{u}{2},q-\frac{u}{2}\big)\,
e^{-i\hbar^{-1}pu}
\end{equation}
with a property
\begin{equation}
\rho(p,p)
=\int\,dq\, W(p,q),
\end{equation}
resembling the classical feature of the distribution function $f(p)=\int F(p,r)dr$.

The second way to define the Wigner function is through the Weyl quantization scheme.

\subsection{Wigner-Weyl symbols}

The Wigner's quasi-distribution function is closely related to the Weyl dequantized 
density matrix.

In the Weyl quantization scheme,
the function on the phase space $A(q,p)$ and the quantum operator corresponding to it $\hat{A}(\hat{q},\hat{p})$ are connected through
their integral Fourier expansions. By the replacement of the coordinate and the momentum by their 
quantum operators $q\longrightarrow\hat{q},\;p\longrightarrow\hat{p}$ in
\begin{equation}
A(q,p)=\frac{1}{(2\pi)^{2}}\int d\sigma\int d\tau\,\widetilde{A}(\sigma,\tau)e^{i(\sigma q+\tau p)},
\end{equation}
one obtains the operator
\begin{equation}
\hat{A}(\hat{q},\hat{p})=
\frac{1}{(2\pi)^{2}}
\int d\sigma\int d\tau\,\widetilde{A}(\sigma,\tau)\,e^{i(\sigma \hat{q}+\tau \hat{p})}.
\end{equation}
Assuming the expansion $e^{\hat O}=1+\hat{O}+\frac{1}{2!}\hat{O}^{2}+\dots$,
and noting that because of the non-commuting $\hat{q}$ and $\hat{p}$ the binomial expansion  is not exactly valid, one observes that by the
Weyl correspondence to a classical  $q^{n}p^{m}$ one associates the symmetrical expression
\begin{equation}
 \frac{1}{2^{n}}\sum_{k=0}^{n}{n \choose k}\, \hat{q}^{n-k}\hat{p}^{m}\hat{q}^{k}=
\frac{1}{2^{m}}\sum_{l=0}^{m}{m \choose l}\, \hat{p}^{m-l}\hat{q}^{n}\hat{p}^{l},
\end{equation}

The inverse Fourier transform closure, yields the appropriate Dirac delta representation.

In the complex phase space
the Weyl quantization scheme  is defined using the  generalized from of 
 the ordinary variables $a=(2\hbar)^{-1/2}\big(q+ip\big),a^{+}=(2\hbar)^{-1/2}\big(q-ip\big)$
differing  in the scaling parameter
\begin{eqnarray}\label{eq:aalam}
&&a=(2\hbar)^{-1/2}\big(\lambda q+i\lambda^{-1}p\big),
\nonumber\\
&&a^{+}=(2\hbar)^{-1/2}\big(\lambda q-i\lambda^{-1}p\big),
\end{eqnarray}
with the same commutation relation $[{\hat a},{\hat a}^{+}]=\hat{1}$.
One also keeps the  Dirac delta representation for the real variables unchanged.
 The Fourier transform of the function of the complex variable $A(\xi)\equiv A(\xi,\xi^\ast)$ 
now reads
\begin{equation}
\widetilde{A}(\alpha)\equiv\widetilde{A}(\alpha,\alpha^\ast)=\frac{1}{\pi}\int\,d^{2}\xi\,A(\xi,\xi^\ast)\,e^{a\xi^\ast-a^\ast\xi},
\end{equation}
with $d^{2}\xi=d{\Re}\xi\,d{\Im}\xi$.
The corresponding delta function closing the transformation, is
using the Dirac delta function parity $\delta(x)=\delta(-x)$, shown to be
\begin{equation}
\delta^{(2)}(\xi)=\delta({\Re}\xi)\,\delta({\Im}\xi)=\frac{1}{\pi^{2}}\int\,d^{2}a\,e^{\,\xi a^\ast-\xi^\ast a}
=\frac{1}{\pi^{2}}\int\,d^{2}a\,e^{\,2{\Im}(\xi a^\ast)}.
\end{equation}

By the Weyl quantizations scheme the function and the operator are related as follows.
With the function of complex variables $A(\xi)$,
one associates the operator $\hat{A}$,
given in terms of the displacement operator, an exponential function 
$\hat{D}(\alpha)=\exp\big(\alpha {\hat a}^{+}-\alpha^\ast {\hat a}\big)$,
by
\begin{equation}\label{eq:WQ2S}
\hat{A}=\frac{1}{\pi}\int\, A(\xi)\hat{D}^{-1}(\xi)d^{2}\xi.
\end{equation}
The function is regained, using $\Tr\!\big[\hat{D}(\alpha)\hat{D}^{-1}(\beta)\big]=\pi \delta^{(2)}(\alpha-\beta)$, by
\begin{equation}\label{eq:DWQ2S}
A(\alpha)=\Tr\!\big[\hat{A}\hat{D}(\alpha)\big].
\end{equation}
This is analogous to the characteristic function
introduced by Moyal in the real phase space
\begin{equation}\label{eq:char}
C(\sigma,\tau)=\Tr\!\big{(}\hat\rho\hat{C}\big{)}=\Tr\!\big{(}\hat\rho\,
e^{\frac{i}{\hbar}(\sigma \hat{q}+\tau \hat{p})}\big{)}.
\end{equation}

It is well known that the  Fourier transformed Wigner function is exactly the characteristic function. 
In the complex phase space,
the Wigner's symbol corresponding to $\hat{X}$, is therefore  a Fourier transform of the
function $X$ reproducing $\hat{X}$ by the Weyl quantization rule
\begin{equation}
W_{\hat{X}}=\frac{1}{\pi}\int\,d^{2}\beta\Tr\Big{(}
\hat{X}\hat{D}(\beta)
\Big{)}e^{\alpha\beta^\ast-\alpha^\ast\beta}.
\end{equation}

\subsubsection{Wigner function as expectation value}

Note that the physical definition of the Wigner function  is a one dimensional integral,
as where the Weyl-Wigner definition is a two-dimensional integral.
However, it reduces to the original form in Lie group Lie algebra correspondence. Either by insertion of the identity resolutions, 
of both coordinate and momentum representations, finalized by the use of the transition amplitude. Or by using the group of translations characteristics in the coordinate representation.

Let us find the characteristic function (\ref{eq:char}), for the density matrix (\ref{eq:den}). Using the Backer-Hausdorf formula one can rewrite the displacement operator
in standard and anti-standard form. In order to find the trace, one inserts the identity resolutions  
in the coordinate and momentum representations
$\hat{1}_{q}=\int dq\ket{q}\bra{q}$ and $\hat{1}_{p}=\int dp\ket{p}\bra{p}$, the first twice. Hence 
\begin{equation}
C(\tau,\sigma)=\frac{1}{2\pi\hbar}\int dp\int dq\int dq^\prime\,\sum_{k}w_{k}\varphi_{k}(q){\varphi}^{\ast}_{k}(q^\prime)e^{\frac{i}{\hbar}p(\tau+q^\prime-q)}
\begin{cases}
e^{\sigma(q-\frac{\tau}{2})},\quad\text{for}\; \hat{p}\hat{q}\\
e^{\sigma(q^\prime+\frac{\tau}{2})},\quad\text{for}\; \hat{q}\hat{p}.
\end{cases}
\end{equation}
Changing the coordinates into $q_\pm=\frac{1}{2}(q\pm q^\prime)$, and integrating over $p$ to obtain
$\delta(\tau-2q_{-})=\frac{1}{2}\delta(q_{-}-\frac{\tau}{2})$ one finds
\begin{equation}
C(\tau,\sigma)=\int dq_{+}\,\sum_{k}w_{k}\,
\varphi_{k}(q_{+}+\frac{\tau}{2})
{\varphi}^{\ast}_{k}(q_{+}-\frac{\tau}{2})\,
e^{\frac{i}{\hbar}\sigma q_{+}}.
\end{equation}
Its Fourier transform is
\begin{equation}
C(\widetilde{q},\widetilde{p})=
\frac{1}{2\pi\hbar}\int d\tau\,\sum_{k}w_{k}\,
\varphi_{k}(\widetilde{q}+\frac{\tau}{2})
{\varphi}^{\ast}_{k}(\widetilde{q}-\frac{\tau}{2})\,
e^{-\frac{i}{\hbar}\tau\widetilde{p}}.
\end{equation}
In this way, the counterpart of ordering, which is a purely quantum physics technique, vanished by the use of the coordinate re-scaling,
in obtaining the classical Wigner function, the Fourier transformed characteristic function.
The sameness of the ordering outcomes  was first observed by Moyal \cite{HOSW}.

A somewhat different
method in calculating Wigner's function, is presented  in \cite{SZ}
and it consists of square rooting  the translation generator $e^{-\frac{i}{\hbar}\tau \hat{p}}$ while calculating in the coordinate representation,
followed by the cyclic permutation within trace. 
The Fourier transform used here is
over the real phase space, in which 
the Fourier transform and its inverse are distinguishable.
Hence,
\begin{eqnarray}
W(q,p)&=&\frac{1}{(2\pi)^{2}}\int d\sigma\int d\tau\,C(\sigma,\tau)\,e^{-i(\sigma q+\tau p)}
\nonumber\\
&=&\frac{1}{(2\pi)^{2}}\int d\sigma\int d\tau\,\Tr\!\big{(}\hat\rho\,
e^{\frac{i}{\hbar}(\sigma \hat{q}+\tau \hat{p})}\big{)}\,e^{-i(\sigma q+\tau p)}
\nonumber\\
&=&\frac{1}{(2\pi)^{2}}\int d\sigma\int d\tau\,\Tr\!\big{(}\hat\rho\,
e^{-\frac{i}{\hbar}(\sigma \hat{q}+\tau \hat{p})}\big{)}\,e^{i(\sigma q+\tau p)}.
\end{eqnarray}
Applying the Backer-Hausdorf formula and 
\begin{equation}
\bra{q}e^{-\frac{i}{\hbar}\tau \hat{p}}\ket{q}=\bra{q}e^{-\frac{i}{2\hbar}\tau \hat{p}}e^{-\frac{i}{2\hbar}\tau \hat{p}}\ket{q}=
\braket{q-\tau/2}{q+\tau/2},
\end{equation} one obtains the standard form of the Wigner function (\ref{eq:Wig}).

To emphasis the definition of the Wigner function, one also uses 
$W(q,p)=\Tr(\hat\rho\Delta(q,p))$
with $\Delta$ being the Wigner's operator \cite{FG}, the Stratonovitch-Weyl operator kernel
\begin{equation}
\Delta(q,p)=\frac{1}{2\pi}
\int du\,
\ket{q-{u}/{2}}
\bra{q+{u}/{2}}
e^{-ipu}.
\end{equation}

In view of the fact that the proposed technique mixes the coordinates from two dual spaces there was a proposal of the 
Weyl-Underhill-Emmrich quantization \cite{PPT},
where one discuses the neighborhoods for the existence of the normal coordinates and generalizes the procedure by distinguishing the 
measures on the curved and flat spaces.

\section{Husimi function}
\cleq

The Wigner function united two descriptions.
It connected the coordinate and momentum representations of the quantum state
described by a density operator,
what is an original interpretation.
Secondly, it is given in terms of the classical counterpart of the quantum state operator,
both being related in a Weyl 
quantization procedure by the displacement operator.

The definition of two other most used quasi-distributions
$Q$ and $P$ is through coherent states.
The coherent states are the most classical quantum states of a harmonic oscillator,
and they can be generated by a classical oscillating current \cite{IQO}.
The modeling of the quantum random walks is due to transition amplitudes between coherent states \cite{QprobQFT}.

The eigenvalues of position and momentum operators,
within Hamiltonian $\hat{H}=\frac{\hat{p}^{2}}{2m}+ U(\hat{x})$ where continuous, and their eigenstates orthogonal but normalized to Dirac delta function
\begin{equation}
\bra{x'}\hat{x}=\bra{x'}x',
\quad
(x'-x)\braket{x'}{x}=0,\quad
\braket{x'}{x}=\delta(x'-x),
\end{equation}
and the relation between coordinate and momentum representations of quantum state given by the Fourier transform
\begin{equation}
\Psi(p,t)=\frac{1}{\sqrt{2\pi\hbar}}\int\psi(q,t)e^{-iqp/\hbar}dq,
\end{equation}
along the path of the free particle with the assigned coordinate $q$.
In particular, in quantum optics the harmonic oscillator with a quadratic potential $U(x)=\frac{1}{2}m\omega^2x^2$ became relevant for describing the motion of the trapped ions and atoms in standing waves cooled to temperatures 
revealing their specific quantum behavior, and also representing each mode  of the quantized electromagnetic field.
A single mode free field  described by the Hamiltonian $\hat{\cal H}=\hbar\omega(\hat{a}^{+}\hat{a}+\frac{1}{2})$.

The ideal approximated state of light in a laser is a coherent state.
The laser provides a stimulated emission of photons in a determined direction induced by an electrical current, with photons being released within an electron relaxation, hence being
related to energy level gaps. Because the laser beam resembles the classical trajectory
the quantum state of photons within has to be the coherent state superposed of photon number states.

The standard definition of coherent states is either through the annihilation or the displacement (doubled translation) operators
\begin{equation}
\hat{a}\ket{\alpha}=\alpha\ket{\alpha},\quad \hat{D}(\alpha)\ket{0}=\ket{\alpha}.
\end{equation}
In general, the coherent states are the infinite sums of the number states, with  numbers representing the determined 
amount of the elementary quanta, which is up to a factor $\alpha$, labeling the states, left unchanged after the annihilation of an elementary quantum \cite{JPG}. The form of the expansion coefficients $\Phi_{n}(\alpha)$ at the coherent states, 
classifies them and determines the "optical phase space" as the image of the map
\begin{equation}
\alpha\rightarrow \xi_\alpha=\sqrt{\bar{n}(\alpha)}e^{i\arg{\alpha}},\quad
\bar{n}(\alpha)=\sum\,n|\Phi_{n}(\alpha)|^{2}.
\end{equation} 

Physically less intuitive is a definition through the displacement operator,
but mathematically this is a key element of quantization.
In definition of coherent states one uses  the representation of Heisenberg-Weyl group
\begin{equation}
e^{i\kappa\hat{1}+\frac{i}{\hbar}\big(
p\hat{Q}-x\hat{P}
\big)},
\end{equation}
representing  the real phase space
in the space of the square integrable complex functions.
With the square integrability prerogative for the probability existence.

The Husimi function is defined for the standard coherent states of the harmonic oscillator,
with $\lambda=(m\omega)^{1/2}$ for its complex phase space variables (\ref{eq:aalam}), by
\begin{equation}\label{eq:H}
Q(q,p)\equiv D_{H}(q,p)=\frac{1}{2\pi h}\bra{\alpha}\hat\rho\ket{\alpha}.
\end{equation}

It is well known that the Husimi quasi-distribution \cite{ADM1,ADD1}, responds to the 
transformations of scale in the real phase space. So that,
under 
a scaling \cite{DL2} $(q,p)\rightarrow(\lambda q,\lambda p)$, the Husimi function $Q(q,p)$ of any
physical state is converted into a function which is also the Husimi function $\lambda^{2}Q(\lambda q,\lambda p)$ of a so called stretched physical state, provided that $|\lambda|^{2}\le 1$.

The coherent states have been generalized.
In use are the squeezed states, the states of radiation fields for which the fluctuation for one quadrature  is reduced from the symmetric quantum limit,
at the cost of enhanced second with no violation of Heisenberg uncertainty principle.
These are applicable in gravitational wave detection \cite{SZ}.
They can be generated by a classical oscillating current in presence of barriers or quadratic displacement potentials.

\subsection{The squeezing procedure}

The non-commuting operators have made possible different ways of
calculating the exponential function, and position of the basic operators (in the normal quantum coordinate operators approximations)
induced the notion of ordering.
For  exponentials of higher orders one needs to incorporate the
Lie algebra isomorphisms in order to define the calculation outcomes.

The squeeze operator, characterized by $\zeta=re^{i\theta}$, defined  by
\begin{equation}
\hat{S}(\zeta)=e^{\frac{1}{2}\big{(} \zeta^\ast\hat{a}^{2}-\zeta\hat{a}^{+2}\big{)}},
\end{equation}
 gives the squeezed coherent state by $\hat{D}(\alpha)\hat{S}(\zeta)\ket{0}=\ket{\alpha,\zeta}$.
Because the algebra of the creation and annihilation operators is $[\hat{a},\hat{a}^{+}]=\hat{1}$,
for their squares, one has
\begin{eqnarray}
&&[\hat{a}^{2},\hat{a}^{+2}]=2\{\hat{a},\hat{a}^{+}\},\nonumber\\
&&[\hat{a}^{2},[\hat{a}^{2},\hat{a}^{+2}]]=8\hat{a}^{2},\nonumber\\
&&[\hat{a}^{+2},[\hat{a}^{2},\hat{a}^{+2}]]=-8\hat{a}^{+2},\nonumber\\
&&[\hat{a}^{+2},[\hat{a}^{2},[\hat{a}^{2},\hat{a}^{+2}]]]
=[\hat{a}^{2},[\hat{a}^{+2},[\hat{a}^{2},\hat{a}^{+2}]]]=-16\{\hat{a},\hat{a}^{+}\},\dots
\end{eqnarray}
and therefore needs the general form of the Baker-Cambell-Hausdorf relation in order to decompose the squeezing operator.
Note that $\{\hat{a},\hat{a}^{+}\}=\frac{2}{\hbar\omega}\hat{\cal H}$.
In matrix linear approximation, one can observe that three matrices closing this algebra share the same Lie algebra
\cite{BCH} with
\begin{eqnarray}
\frac{1}{4}\{\hat{a},\hat{a}^{+}\}&\longleftrightarrow& K_{0}=-\frac{1}{2}\sigma_{z}=\frac{1}{2}\begin{bmatrix}
-1 & 0\\
0 & 1
\end{bmatrix},
\nonumber\\
\frac{1}{2}\hat{a}^{2}&\longleftrightarrow& 
K_{-}=\sigma_{+}=\begin{bmatrix}
0 & 1\\
0 & 0
\end{bmatrix},
\nonumber\\
 \frac{1}{2}\hat{a}^{+2}&\longleftrightarrow&
K_{+}=\sigma_{-}=\begin{bmatrix}
0 & 0\\
-1 & 0
\end{bmatrix}.
\end{eqnarray}
Their exponential functions  are simply
\begin{equation}
e^{cK_{0}}=\begin{bmatrix}
e^{-{c}/{2}} & 0\\
0 & e^{{c}/{2}}
\end{bmatrix},
\end{equation}
together with
\begin{equation}
e^{cK_{-}}=\hat{1}+cK_{-}=\begin{bmatrix}
1 & c\\
0 & 1
\end{bmatrix},\qquad
e^{cK_{+}}=\hat{1}+cK_{+}=\begin{bmatrix}
1 & 0\\
-c & 1
\end{bmatrix}.
\end{equation}
Therefore, the decomposition of the squeezing operator can be obtained by
equating the product of three matrices exponentials $e^{aK_{+}}e^{cK_{0}}e^{bK_{-}}$ and the representation of the squeezing operator 
$e^{-\zeta K_{+}+\zeta^\ast K_{-}}$
\begin{eqnarray}
\exp\Big{\{}
\begin{bmatrix}
0 & \zeta^\ast\\
\zeta & 0
\end{bmatrix}
\Big{\}}&=&
\sum_{k=0}^\infty\frac{1}{(2k)!}\begin{bmatrix}
|\zeta|^{2k} &0\\
0 & |\zeta|^{2k}
\end{bmatrix}
+\sum_{k=0}^\infty\frac{1}{(2k+1)!}\begin{bmatrix}
0 & \zeta^\ast|\zeta|^{2k}\\
\zeta|\zeta|^{2k} & 0
\end{bmatrix}
\nonumber\\
&=&
\begin{bmatrix}
\cosh{|\zeta|} & \frac{\zeta^\ast}{|\zeta|}\sinh{|\zeta|}\\
\frac{\zeta}{|\zeta|}\sinh{|\zeta|} & \cosh{|\zeta|}
\end{bmatrix}.
\end{eqnarray}
One finds normally ordered squeezing operator decomposition
\begin{equation}
S(\zeta)=e^{-\frac{1}{2}e^{i\theta}\tanh{|\zeta|}\hat{a}^{+2}}
e^{-\ln(\cosh{|\zeta|})(\frac{1}{2}+\hat{a}^{+}\hat{a})}
e^{\frac{1}{2}e^{-i\theta}\tanh{|\zeta|}\hat{a}^{2}}.
\end{equation}
Hence, the squeezed vacuum state is given by
$\ket{0,\zeta}=(\cosh{|\zeta|})^{-1/2}e^{-\frac{1}{2}e^{i\theta}\tanh{|\zeta|}\hat{a}^{+2}}\ket{0}$.

There is also a non-unitary approach \cite{AWSS}, by which 
the squeezed vacuum states are
\begin{equation}
\ket{0,0;\zeta}\equiv(1+\zeta\zeta^\ast)^{1/4}\exp(-\frac{1}{2}\zeta a^{2+})\ket{0}
\end{equation}
 and they are eigenstates of the operator $a+\zeta a^{+}$ with the zero eigenvalue
$(a+\zeta a^{+})\ket{0,0;\zeta}=0,$ and normalized for $\braket{0,0;-\zeta}{0,0;\zeta}=1$.

It is known the Husimi function is the Wigner function, smoothed by the "course graining" function
\begin{equation}
Q(q,p,\lambda)=\int d\tilde{q}\int d\tilde{p}\,W(\tilde{q},\tilde{p})\,
e^{-\lambda(\tilde{q}-q)^{2}-\frac{1}{\lambda}(\tilde{p}-p)^{2}}.
\end{equation}
This  definition of the Husimi function  is here given through the generalized coherent state
$\ket{q,p;\Phi}=\hat{U}(q,p)\ket{\Phi}$.
As usual $\hat{U}(q,p)$ is a displacement operator.
The smoothed Wigner function reads
\begin{equation}
D(q,p)=\frac{1}{2\pi}\bra{q,p;\Phi}\hat\rho\ket{q,p;\Phi},
\end{equation}
with the parameter $\Phi$ representing the fiducial vector, characterizing the quantization method \cite{QprobQFT}.
The discussion on smoothed Wigner function can be found in \cite{DL}.
In \cite{FG} this transition
is interpreted as the the pure squeezed state expectation value of the density matrix.

The uncertainty relations for all these types of generalized coherent states, and the correlated states can be found in \cite{VAnd}.
They are characterized by a large uncertainty which for some enlarges the probability for the tunneling effect.
The classical coherent and  squeezed states have been generalized to nonlinear coherent states used to describe the quantum systems beyond the linear harmonic oscillator,
which induces the generalization of the coherent states related distributions and their mutual bonds \cite{Popov}.

\section{Glauber-Sudarshan function}
\cleq

The Glauber-Sudarshan quasi-distribution function \cite{JPG}  actually represents the quantization rule in terms of the coherent states.
The definition is implicit, and it is a so called diagonal form \cite{PLP}
\begin{equation}
\hat{\rho}=\frac{1}{\pi}\int\,d^{2}\beta\,P(\beta)\ket{\beta}\bra{\beta}.
\end{equation}
The definition follows from the identity resolution and the fact 
$\ket{\alpha}\bra{\alpha}=(\ket{\alpha}\bra{\alpha})^{2}$.
Obviously, the Husimi function is a  weighted   Glauber-Sudarshan quasi-probability,
because  
$|\braket{\beta}{\alpha}|^{2}=e^{-|\alpha-\beta|^{2}}$.
It is well known all three  quasi-distributions, $W,Q,P$, are connected by Weierstrass transform \cite{Polj}, smoothing out (also refereed to as an averaging process),
with $P$ being the  base of the smoothing  chain $Q\leftarrow W\leftarrow P$.

Mehta was first to determine the explicit form of the Glauber-Sudarshan quasi-di\-stri\-bu\-tion \cite{IQO},
by calculating
\begin{equation}
\bra{-\upsilon}\hat{\rho}\ket{\upsilon}=\frac{1}{\pi}e^{-|\upsilon|^{2}}\int\,d\alpha^{2}\,P(\alpha)e^{-|\alpha|^{2}}
e^{\alpha^\ast \upsilon-\alpha \upsilon^\ast},
\end{equation}
obtaining the $P$-distribution as the weighted Fourier transform
\begin{equation}
\widetilde{P(\alpha)e^{-|\alpha|^{2}}}=e^{|\upsilon|^{2}}\bra{-\upsilon}\hat{\rho}\ket{\upsilon}.
\end{equation}
 The optical equivalence theorem gives the expectation value of the normally ordered function by $\langle
G(a,a^{+})\rangle=\frac{1}{\pi}\int\,d^{2}\alpha P(\alpha)G(\alpha,\alpha^\ast)$.
Hence, there is a definition of the Glauber-Sudarshan P -- function in form of the  expectation value of the 
normally ordered delta function
\begin{equation}
P(q,q^\prime)=\Tr\Big{(}
\hat{\rho}\delta(\alpha^\ast-\hat{a}^{+})
\delta(\alpha-\hat{a})
\Big{)},
\end{equation}
with
\begin{equation}
\delta(\alpha^\ast-\hat{a}^{+})\delta(\alpha-\hat{a})
=\frac{1}{\pi}
\int\,dc^{2}\,
e^{ic(\alpha^\ast-\hat{a}^{+})}
e^{ic^\ast(\alpha-\hat{a})}.
\end{equation}

\section{Cohen quasi-distributions}
\cleq

Even though the $W,Q,P$ distributions are seen to be connected to
differently ordered delta functions acting on operators, 
the problem is sensitive since the
ordering is lost after the application of operators  is done, possibly finding the new position 
according to the exact matching. To emphasis these variations within the expectation value integral formula Cohen introduced the large class of quasi-distributions.
The Cohen quasi-distributions are the result of combining properties of the transition amplitudes between the
eigenstates of the two operators chosen for the description. The main three being:
amplitudes are invertible, they close on 
Dirac delta and delta by definition satisfies
 $\int dy\delta(y-x)=1$, $\int d^{2}\beta\delta^{(2)}(\beta-\alpha)=1$.

The fluidity between representations in  the Dirac notation,
is demonstrated in paper \cite{BBC}
by considering the  expectation value of the 
arbitrary operator,
$\langle \hat{O} \rangle=\Tr[\hat{O}\hat{\rho}]$, and the operation of unity insertion
\begin{eqnarray}
\int dq_{1}\;\bra{q_{1}}\hat{O}\hat{\rho}\ket{q_{1}}&=&
\int dq_{1}\;\bra{q_{1}}\hat{1}_{p_{1}}\hat{O}\hat{1}_{p_{2}}\hat{1}_{q_{2}}\hat{\rho}\ket{q_{1}}
\nonumber\\
&=&\int dq_{1}\!\int dp_{1}\!\int dp_{2}\!\int dq_{2}\;\bra{p_{1}}\hat{O}\ket{p_{2}}\braket{p_{2}}{q_{2}}\bra{q_{2}}\hat{\rho}\ket{q_{1}}
\braket{q_{1}}{p_{1}}
\nonumber\\
&=&\int dp_{1}\;\bra{p_{1}}\hat{O}\hat{\rho}\ket{p_{1}}.
\end{eqnarray}
Within this formula one will straightforwardly obtain the Wigner symbols for operators $\hat{O}$ and $\hat{\rho}$,
the Wigner function exactly for the second, by simple change of variables:
to difference and mean of both phase space variables, marked by $\bar{q},\bar{p}$ and $q,p$ respectively
\begin{eqnarray}
&&\langle \hat{O} \rangle
=\int dq_{1}\!\int dp_{1}\!\int dp_{2}\!\int dq_{2}\;\bra{p_{1}}\hat{O}\ket{p_{2}}\braket{p_{2}}{q_{2}}\bra{q_{2}}\hat{\rho}\ket{q_{1}}
\braket{q_{1}}{p_{1}}
\\
&&=\int\int dqdp\;\frac{1}{2\pi\hbar}\int d\bar{p}\,\bra{p-\frac{\bar{p}}{2}}\hat{O}\ket{p+\frac{\bar{p}}{2}}e^{-\frac{i}{\hbar}q\bar{p}}
\;\int d\bar{q}\,\bra{q+\frac{\bar{q}}{2}}\hat{\rho}\ket{q-\frac{\bar{q}}{2}}e^{-\frac{i}{\hbar}p\bar{q}}.
\nonumber
\end{eqnarray}
Note that inserting unities in $p$-representation around $\hat{\rho}$ in Wigner function $W(q,p)$,
it can 
for change of momentum $\bar{p}=p_{2}-p_{1}$, and after the application of
$\frac{1}{2\pi\hbar}\int d\bar{q}e^{\frac{i}{\hbar}\bar{q}(p_{1}-p+\frac{\bar{p}}{2})}=\delta(p_{1}-p+\frac{\bar{p}}{2})$
become
\begin{equation}
W(q,p)=
\int d\bar{q}\,\bra{q+\frac{\bar{q}}{2}}\hat{1}_{p_{1}}\hat{\rho}\hat{1}_{p_{2}}\ket{q-\frac{\bar{q}}{2}}e^{-\frac{i}{\hbar}p\bar{q}}
=\int d\bar{p}\,\bra{p-\frac{\bar{p}}{2}}\hat{\rho}\ket{p+\frac{\bar{p}}{2}}e^{-\frac{i}{\hbar}q\bar{p}}.
\end{equation}
Hence 
\begin{equation}
\langle \hat{O}\rangle=\int dq\int \frac{dp}{2\pi\hbar}\;W_{O}(q,p)W(q,p).
\end{equation}
The Wigner's symbol is of the same form for the density matrix and an arbitrary operator, which in a quantum measurement has a role of 
placing the system into one of its eigen-states $\hat{O}\ket{\psi}\rightarrow\ket{\psi_{i}}$.
Hence, not only the ordering is indistinguishable for the Wigner symbol.

Analogously, in the general case of two operators with given eigenvectors $\ket\alpha,\ket\beta$, values $\alpha,\beta$ and transition amplitudes $\braket{\alpha}{\beta}$, one can calculate the 
expectation value of the arbitrary operator by
\begin{equation}
\langle \hat{O} \rangle=\int d\alpha^\prime\int d\alpha^{\prime\prime}{\cal A}^\ast(\alpha^\prime){\cal A}(\alpha^{\prime\prime})\delta(\alpha^{\prime}-\alpha^{\prime\prime})gd\alpha^{\prime}d\alpha^{\prime\prime}.
\end{equation}
Through delta and transition amplitudes one incorporates the other operator and again through delta and its closure one incorporates the second pair of
eigenvalues.

Looking at the product of two delta functions
$\delta(a-\bar{a})\delta(b-\bar{b})$,
one can break and mix their definitions by applying the 
equalities they imply.
First, one can just rearrange the variables joining pairs belonging to different arguments,
or further more  perform the scaling of the dumb variables
to finally obtain  various representations:
\begin{eqnarray}
&&\frac{1}{(2\pi)^{2}}\int d\sigma\!\int d\tau\, e^{i\sigma(a-\bar{a})}e^{i\tau(b-\bar{b})}
\nonumber\\
&=&\frac{1}{(\pi\hbar)^{2}}\int da^\prime\!\int db^\prime
\, e^{-2i(a^\prime-a)(b^\prime-b)/\hbar}
e^{2i(a^\prime-\bar{a})(b^\prime-\bar{b})/\hbar}
\\
&=&\frac{1}{(2\pi)^{4}}\int da^\prime\!\int db^\prime
\int d\theta\!\int d\theta^\prime\int d\tau\!\int d\tau^\prime\,
\frac{\Phi(\theta,\tau)e^{i\theta(\bar{a}-a^\prime)}e^{i\tau\theta\hbar/2}e^{i\tau(\bar{b}-b^\prime)}}{\Phi(\theta^\prime,\tau^\prime)e^{i\theta^\prime(a-a^\prime)}e^{i\tau^\prime\theta^\prime\hbar/2}e^{i\tau^\prime(b-b^\prime)}}\,.
\nonumber
\end{eqnarray}
The first operation produces the Wigner quasi-distribution while the second gives the large class of Cohen quasi-distributions characterized by $\Phi$
\begin{equation}
\frac{1}{(2\pi)^{2}}\int da^\prime db^\prime da^{\prime\prime}d\theta d\tau\,\Phi(\theta,\tau)
{\cal A}^\ast(a^\prime){\cal A}(a^{\prime\prime})
T(\bar{b},a^{\prime\prime})T^{+}(a^{\prime},\bar{b})
e^{i\theta(\bar{a}-a^\prime)}e^{i\tau\theta\hbar/2}e^{i\tau(\bar{b}-{b}^\prime)},
\end{equation}
with $T$ being the transition amplitude.

It is interesting, the insertion of the displacement operator unitary condition $DD^{+}=1$, together with its simple action on the creation and annihilation operators $D^{+}aD=a+\alpha$ and $D^{+}a^{+}D=a^{+}+\alpha^\ast$
induces displacement of states and "re-scaling" of operator
for example making transition from
  the squeezed coherent state to the squeezed vacuum expectation value
$\bra{\zeta,\alpha}a^{+k}a^{k}\ket{\xi,\alpha}\longrightarrow
\bra{\zeta}(a^{+}+\alpha^\ast)^{k}(a+\alpha)^{k}\ket{\xi}$ \cite{AA}.

\section{Electromagnetic field in the amplifier}
\cleq

Let us now as an example consider  a system of $N$ two-level atoms,
where each atom is coupled to a single mode resonant radiation field,
whose interaction is described by the Jaynes-Cummings model \cite{VAnd,Drake78,Drake79}.
More precisely we consider a linear quantum amplifier model, described in \cite{MW},
where $N_1$ out of $N$ atoms are in the excited state,
while the rest of $N_0$ atoms are in the ground state, with $N_0<N_1$.
The occupation numbers are kept constant in time.
The master equation for the electromagnetic field density operator is given,
in the Born-Markov approximation by 
\cite{MW,AT}
\begin{eqnarray}   
\label{V3}
\frac {\partial\hat\rho}{\partial t}=-\gamma N_1(\hat a\hat a^{+}\hat\rho-2\hat a^{+}\hat \rho \hat a+\hat\rho \hat a\hat a^{+})-
\gamma N_0(\hat a^{+} \hat a\hat \rho-2\hat a\hat \rho \hat a^{+}+\hat \rho \hat a^{+} \hat a),
 \end{eqnarray}
with
$\gamma$ being the amplification coefficient. The interaction is in the rotating wave 
approximation.

One can using the definition of the Husimi function, obtain the ordinary differential equation for the Husimi function,
and obtain the Husimi function for the amplified state  as in \cite{AT}:
 \begin{eqnarray}   
\label{V4}
Q(\alpha,t)=\frac{1}{\pi m}\int d^2 \beta\,Q(\beta)\,e^{-\frac{|\alpha-\beta G|^2}{m}},
 \end{eqnarray}
with $G(t)=e^{2(N_1-N_0)\gamma t}$ and $m=\frac{N_0}{N_1-N_0}(G^2-1)$.
 In \cite{ADM1,ADD1} we considered the case  where all atoms are in excited state so that $N_0=0$ and consequently $m=0$. Then, the expression for the Husimi function of the amplified state  is much simpler
\begin{eqnarray}   
\label{V6}
 Q_{out}(\alpha,t)=\frac1{G^2}Q_{in}(\alpha/G)=
\bra{\alpha/G}\hat\rho_{in}\ket{\alpha/G}.
\end{eqnarray}
Obviously, by considering the  dynamics of physical processes one encounters their relationship with both
the  transformations of the phase space and the functions used to describe the process itself.
Depending on the occupation numbers of the content(media) which is here kept constant but generally fluctuating,
one deals with transformation of the phase-space being globally re-scaled or by over all phase-space function course graining.

 Let us now find the approximation for the quantum amplification of the arbitrary operator.
We will consider the bounded operator,  expanded in the normally, anti-normally and symmetrically ordered power series
\begin{equation}
\hat{F}=\sum_{_{ N,M=0}}^\infty\, c_{_{ N,M}}\,(\hat{a}^{+})^{N}\hat{a}^{M}=
\sum_{_{ N,M=0}}^\infty\, d_{_{N,M}}\,\hat{a}^{M}(\hat{a}^{+})^{N}
=\sum_{_{ N,M=0}}^\infty\, e_{_{N,M}}\,\{\hat{a}^{M}(\hat{a}^{+})^{N}\},
\end{equation}
with $\{\}$ marking the symmetric product.
Following \cite{CG}, one can
impose the restrictions on the power series coefficients, using
the recurrent integral expansion,
obtained combining 
the correspondence rules in the complex phase space (\ref{eq:WQ2S}) and (\ref{eq:DWQ2S})
\begin{equation}
\hat{F}=\frac{1}{\pi}\int\Tr(\hat{F}\hat{D}(\xi))\hat{D}^{-1}(\xi)d^{2}\xi.
\end{equation}

After the quantum amplification, three Husimi functions become
\begin{eqnarray}\label{eq:Q3}
Q(\alpha,t)&=&\frac{1}{\pi m}\int d^2 \beta\,\sum_{_{ N,M=0}}^\infty\, c_{_{ N,M}}\beta^{\ast N}\beta^{M}
\,e^{-\frac{|\alpha-\beta G|^2}{m}},
\nonumber\\
Q(\alpha,t)&=&\frac{1}{\pi m}\int d^2 \beta\, \sum_{_{ N,M=0}}^\infty\,d_{_{ N,M}}\beta^{M}\beta^{\ast N}
\,e^{-\frac{|\alpha-\beta G|^2}{m}},
\nonumber\\
Q(\alpha,t)&=&\frac{1}{\pi m}\int d^2 \beta\,\sum_{_{ N,M=0}}^\infty\, \frac{e_{_{ N,M}}}{1+M}\sum_{n=0}^{M}\beta^{M-n}\beta^{\ast N}\beta^{n}
\,e^{-\frac{|\alpha-\beta G|^2}{m}}.
\end{eqnarray}
The base of all three is obviously the same integral. Let us calculate it explicitly.

\subsection{Husimi base integrals}
Let us calculate the integral
\begin{equation}
 I_{_{ N,M}}=\int d^2 \beta\,\beta^{M}\beta^{\ast N}
\,e^{-\frac{|\alpha-\beta G|^2}{m}}.
\end{equation}
Taking the change of variables $z=\frac{\alpha-\beta G}{\sqrt{m}}$, one has $\beta=\frac{\alpha-\sqrt{m}z}{G}$
and therefore obtains
\begin{eqnarray}
I_{_{ N,M}}&=&\frac{m}{G^{M+N+2}}\int d^{2}z\,
\big{(}
{\alpha-\sqrt{m}z}
\big{)}^{M}
\big{(}
{\alpha-\sqrt{m}z}
\big{)}^{\ast N}
\,e^{-{|z|^2}}
\nonumber\\
&=&\frac{m}{G^{M+N+2}}\int d^{2}z\,
\sum_{i=0}^{M}{M\choose i}\alpha^{i}z^{M-i}
\sum_{j=0}^{N}{N\choose j}\alpha^{\ast j}(z^{\ast})^{N-j}
\nonumber\\
&\cdot&(-\sqrt{m})^{M-i+N-j}\,e^{-{|z|^2}}.
\end{eqnarray}
Changing the variables to polar coordinates, the integral becomes
\begin{eqnarray}
I_{_{ N,M}}&=&
\frac{m}{G^{M+N+2}}
\int_{0}^\infty \rho d\rho \int_{0}^{2\pi}d\varphi\,
\sum_{i=0}^{M}{M\choose i}\alpha^{i}
\sum_{j=0}^{N}{N\choose j}\alpha^{\ast j}
\nonumber\\
&\cdot&
\rho^{M-i+N-j} e^{i\varphi(M-i-(N-j))}
(-\sqrt{m})^{M-i+N-j}\,e^{-{|z|^2}}.
\end{eqnarray}
The integration over $\varphi$ yields $\int d\varphi\,e^{i\varphi(M-i-(N-j))}=0$ for $M-i\neq N-j$. Hence
\begin{eqnarray}\label{eq:IMN}
I_{_{ N,M}}&=&
\frac{m}{G^{M+N+2}}
\int_{0}^\infty \rho d\rho 2\pi\,
\sum_{j=max(0,N-M)}^{N}{M\choose M-N+j}\alpha^{M-N+j}{N\choose j}\alpha^{\ast j}
\nonumber\\
&\cdot&
\rho^{2(N-j)} 
m^{N-j}\,e^{-{\rho^2}}.
\end{eqnarray}
Let us now consider the integration over $\rho$, and mark
\begin{equation}
{\cal I}_{2(N-j)+1}=\int_{0}^\infty \rho d\rho\,\rho^{2(N-j)}\,e^{-{\rho^2}}=
\int_{0}^\infty d\rho\,\rho^{2(N-j)+1}\,e^{-{\rho^2}}.
\end{equation} 
The first integral is just
\begin{equation}
{\cal I}_{1}=\int_{0}^\infty d\rho\,\rho^{1}\,e^{-{\rho^2}}=
-\frac{1}{2}\int_{0}^\infty d(-\rho^{2})\,e^{-{\rho^2}}=-\frac{1}{2}e^{-{\rho^2}}\Big{|}_{0}^\infty=1/2.
\end{equation}
There is a recurrent relation, between consecutive integrals, obtained by the partial integration
\begin{equation}
{\cal I}_{2n+1}=\int_{0}^\infty d\rho\,\rho^{2n+1}\,e^{-{\rho^2}}=
-\frac{1}{2}\rho^{2n}e^{-{\rho^2}}\Big{|}_{0}^\infty
+n\int_{0}^\infty d\rho\,\rho^{2n-1}\,e^{-{\rho^2}}=n{\cal I}_{2n-1}.
\end{equation}
Therefore
${\cal I}_{3}=1{\cal I}_{1}=1/2$, ${\cal I}_{5}=2{\cal I}_{3}=1$, ${\cal I}_{7}=3{\cal I}_{1}=3$,
${\cal I}_{9}=4{\cal I}_{7}=4\cdot 3=12$ so that one can conclude
${\cal I}_{2n+1}=n!/2$ and confirm that
\begin{equation}
{\cal I}_{2n+3}=(n+1){\cal I}_{2n+1}=(n+1)n!/2=(n+1)!/2.
\end{equation}
Into the  equation (\ref{eq:IMN}), we therefore substitute ${\cal I}_{2(N-j)+1}=(N-j)!/2$, and conclude
\begin{eqnarray}\label{eq:IMNfin}
I_{_{ N,M}}&=&
\frac{2\pi m}{G^{M+N+2}}
\sum_{j=max(0,N-M)}^{N}\frac{(N-j)!}{2}{M\choose M-N+j}\alpha^{M-N+j}{N\choose j}\alpha^{\ast j}
m^{N-j}
\nonumber\\
&=&\frac{\pi m}{G^{M+N+2}}
\sum_{j=max(0,N-M)}^{N}\frac{M!{N\choose j}}{(M-N+j)!}\alpha^{M-N+j}\alpha^{\ast j}
m^{N-j}.
\end{eqnarray}

The obtained functions are the building blocks of the boundary ordering valued quasi-distributions (\ref{eq:Q3}), used for the evaluation of the actual statistics.


\section{Conclusion}

The ordering is highly sensitive problem.
It emerges with non-commutativity, which itself originates from the correspondence rules between the classical and quantum phase spaces,
and the restrictions for the quantum measurements. The essential difference being the absolute determination of position and momentum,
and the probabilistic quantum nature described by the quantum state. Which at first described the quantum on the subatomic level,
but with lasers obtained its the most classical stimulated manifestation. 

The same as the quantum is the joint measure of energy, in discussion of distances  the accordance of the matrices is needed.
This is what defines the string quantization, the conservation of the constraints for the consistent  distance measures on the world-sheet and the space-time.
In string theory, restrictions on the plane wave normal coordinates is made by discussion of the boundary conditions and in addition of their-own conservation.
 
The defining algebraic properties cause the separate problems in the actual calculations, which extend with the powers of the non-commuting operators
used to describe the concrete phenomenon. Because their algebra is postulated, its incorporation into the quantum theory is actually done by means of 
discussion of the different possible orderings. Here, closely related to the expansion of the displacement operator. 

The quantum state is in the phase-space formulation of quantum mechanics represented by the quasi-probability function of the phase-space variables.
The most used being Wigner, Husimi and Glauber-Sudarshan quasi-probabilities. Although they can be related to 
specific orderings through Dirac delta product  representations, it can be algebraically shown that Wigner function is insensible to ordering, using coordinate re-scaling.
The connection between these quasi-distributions is twofold, first through representations of the same state of nature, second through the connection of 
quantization techniques, embedding quanta into the non-local coherent beam. 

In string theory measure and time embedding of the world-sheet into the space-time
involve the separation of space-time coordinates into light-cone and transverse directions,
making one of the space-time coordinates  
entirely determinable by the transverse directions.
In the case of the string with Neumann boundary conditions these directions represent 
the discrete frequency oscillations.

The coherent state being the best connection between the classical and quantum worlds, singles out  the Husimi quasi-distribution.
In this paper, we have explicitly calculated 
the Husimi quasi-probability function for the 
generically ordered operators fulfilling the evolution equation for the two level quantum amplifier.

The analysis of whether the quantum state can be described by two, coordinate and momentum distributions,
sometimes called marginal,
question set by Pauli \cite{PAU}, evolved into a quantum tomography scheme \cite{MMT}.
Starting with Wigner function $W(q,p)$, representing symmetrical ordering,
and more significantly by Wigner a quantum mechanical state on a phase space,
 by an invertible transformation one defines 
a two parameter function $w(X,\mu,\nu)$,
the symplectic tomogram of the quantum state. It represents the family of quasi-distributions for $X=\mu q+\nu p$
being the coordinate in the rotated phase space system and  describes the set of measurements of the non-commuting 
observables.

There is a structural resemblance to representing the physically equivalent T-dual string theories
on the unique phase space in a symplectic T-duality. 
Here, the canonical Hamiltonian, and two symplectic  representations of T-dual theories 
are related by 
\begin{eqnarray}\notag
{\cal H}&=&\frac{1}{2\kappa} (X^T)^M \begin{bmatrix}
G^E_{\mu \nu} & - 2B_{\mu \rho} (G^{-1})^{\rho \nu} \\
2(G^{-1})^{\mu \rho} B_{\rho \nu} & (G^{-1})^{\mu \nu}
\end{bmatrix}
\ X^N
\nonumber\\&=&\frac{1}{2 \kappa}  (\hat{X}^T)^M\ \begin{bmatrix}
G_{\mu \nu} & 0 \\
0 & (G^{-1})^{\mu \nu}
\end{bmatrix}\ \hat{X}^N
\nonumber\\
&=&\frac{1}{2 \kappa}  (\tilde{X}^{T})^M \begin{bmatrix}
G^E_{\mu \nu} & 0 \\
0 & (G_E^{-1})^{\mu \nu} 
\end{bmatrix}
 \tilde{X}^N.\nonumber
\end{eqnarray}
The original is depending on the generalized metric given in terms of metric and Kalb-Ramond field
with phase space coordinates obeying $\{x^\mu,\pi_\nu\}=\delta^\mu_\nu$,
while the two T-dual theories are represented as topology free and are given in terms of the  non-canonical variables, the currents $k^\mu,i_\mu$
$$X^M=\begin{bmatrix}
\kappa x^{\prime \mu} \\
\pi_\mu \\
\end{bmatrix},
\quad \hat{X}^M=
\begin{bmatrix}
\kappa x^{\prime \mu} \\
i_\mu
\end{bmatrix}
=(e^{\hat{B}})^M_{\ N}\ X^N,
\quad
\tilde{X}^M=
\begin{bmatrix}
k^\mu \\
\pi_\mu 
\end{bmatrix}=(e^{\hat{\theta}})^M_{\ N} X^N $$
which generate a generalized Poisson structure
\begin{eqnarray}
&&\{ i_\mu (\sigma), i_\nu (\bar{\sigma}) \} = - 2\kappa B_{\mu \nu \rho} x^{\prime \rho} \delta (\sigma - \bar{\sigma}),\nonumber\\
&&\{ k^\mu (\sigma), k^{\nu} (\bar{\sigma}) \} = -\kappa Q_\rho^{\ \mu \nu} k^\rho \delta(\sigma - \bar{\sigma}) - \kappa^2 R^{\mu \nu \rho} \pi_\rho \delta (\sigma - \bar{\sigma}).
\end{eqnarray}
The  obtained structural constants are the fluxes of the string's background.
The anti-symmetric Kalb-Ramond field produces  the $H$-flux: $B_{\mu \nu \rho}= \partial_\mu B_{\nu \rho} + \partial_\nu B_{\rho \mu} + \partial_\rho B_{\mu \nu}$ which is its field strength,
responsible for non-associativity of the star product in a deformed quantization procedure.
The T-dual Kalb-Ramond field $\theta^{\mu\nu}$ being also the non-commutativity parameter creates
the $Q$-flux: $Q_{\rho}^{\ \mu \nu} = \partial_\rho \theta^{\mu \nu}$ determining the form of the non-commutativity (possibly nonlocal, T-dual, hence nongeometrical)
and 
$R$-flux: $R^{\mu \nu \rho} = \theta^{\mu \sigma} \partial_\sigma \theta^{\nu \rho} + \theta^{\nu \sigma} \partial_\sigma \theta^{\rho \mu} + \theta^{\rho \sigma} \partial_\sigma \theta^{\mu \nu}$ having a form of the generalized Poisson bracket Jacobiator is influencing the algebraic properties and volume uncertainty relation.

It is interesting the symplectic tomogram, which is proclaimed 
the most realistic probability distribution introduced,
is defined through the Wigner function using the  integral Radon transform,
which however is not applicable to Dirac delta functions being the Wigner functions for the
de Broglie plane wave.
The resolution of the problem is found in using  the delta function in its restricted form \cite{MM}.


\end{document}